\documentclass[sn-mathphys,Numbered,iicol]{sn-jnl}

\usepackage{graphicx}%
\usepackage{multirow}%
\usepackage{amsmath,amssymb,amsfonts}%
\usepackage{amsthm}%
\usepackage{mathrsfs}%
\usepackage[title]{appendix}%
\usepackage{xcolor}%
\usepackage{textcomp}%
\usepackage{manyfoot}%
\usepackage{booktabs}%
\usepackage{algorithm}%
\usepackage{algorithmicx}%
\usepackage{algpseudocode}%
\usepackage{listings}%
\usepackage{siunitx}
\usepackage{subfig}

\begin{document}

\title[A bias using the ages of OAO to the Hubble tension]{A bias using the ages of the oldest astrophysical objects to address the Hubble tension}


\author*[1,2]{\fnm{André A.} \sur{Costa}}\email{andrecosta@yzu.edu.cn}

\author[2]{\fnm{Zelin} \sur{Ren}}

\author[2]{\fnm{Zhixiang} \sur{Yin}}

\affil*[1]{\orgdiv{College of Physics}, \orgname{Nanjing University of Aeronautics and Astronautics}, \orgaddress{\city{Nanjing} \postcode{211106}, \country{China}}}

\affil[2]{\orgdiv{Center for Gravitation and Cosmology, College of Physical Science and Technology}, \orgname{Yangzhou University}, \orgaddress{\city{Yangzhou} \postcode{225009}, \country{China}}}


\abstract{Recently different cosmological measurements have shown a tension in the value of the Hubble constant, $H_0$. Assuming the $\Lambda$CDM model, the {\it Planck} satellite mission has inferred the Hubble constant from the cosmic microwave background (CMB) anisotropies to be $H_0 = 67.4 \pm 0.5 \, \rm{km \, s^{-1} \, Mpc^{-1}}$. On the other hand, low redshift measurements such as those using Cepheid variables and supernovae Type Ia (SNIa) have obtained a significantly larger value. For instance, Riess et al. reported $H_0 = 73.04 \pm 1.04 \, \rm{km \,  s^{-1} \, Mpc^{-1}}$, which is $5\sigma$ apart of the prediction from {\it Planck} observations. This tension is a major problem in cosmology nowadays, and it is not clear yet if it comes from systematic effects or new physics. The use of new methods to infer the Hubble constant is therefore essential to shed light on this matter. In this paper, we discuss using the ages of the oldest astrophysical objects (OAO) to probe the Hubble tension. We show that, although this data can provide additional information, the method can also artificially introduce a tension. Reanalyzing the ages of 114 OAO, we obtain that the constraint in the Hubble constant goes from slightly disfavoring local measurements to favoring them. }




\maketitle


\section{Introduction}
\label{Sec.1}
Currently several measurements have shed light in our understanding of the cosmos. One of such measurements, the cosmic microwave background (CMB) provides information about the early universe. At that time, protons and electrons formed the first neutral atoms and decoupled from photons, which traveled freely until we observe them nowadays. The CMB photons possesses a simple physical explanation and carry information about the background and perturbations, which can be modeled linearly, at those early stages. Several experiments have been dedicated to measure those photons, the latest and most precise one was accomplished by the {\it Planck} satellite \cite{Planck:2018nkj}.

In order to make predictions about the CMB photons coming from the distant past, we have to assume a specific model. Therefore, the comparison with data implies constraints on late-time cosmological parameters, in particular, the Hubble constant, $H_0$. Assuming the $\Lambda$CDM model, the {\it Planck} satellite mission have constrained the value of $H_0$ very precisely as $H_0 = 67.4 \pm 0.5 \, \rm{km \, s^{-1} \, Mpc^{-1}}$ \cite{Planck:2018vyg}.

In addition, we also have observations from the light of stars, galaxies, and other astrophysical objects at much lower redshifts. Their distance and clustering provide information about our Universe at more recent times. In particular, observations from the {\it Hubble Space Telescope} (HST) of Cepheid variables in the host galaxies of 42 Type Ia supernovae (SNIa) were used to calibrate the Hubble constant \cite{Riess:2021jrx}. Their baseline result was $H_0 = 73.04 \pm 1.04 \, \rm{km \,  s^{-1} \, Mpc^{-1}}$, which is $5\sigma$ apart of the prediction from {\it Planck} observations under the $\Lambda$CDM model.

This tension in the Hubble constant value is a major problem in cosmology today. It is not clear yet, if the tension comes from some systematic effect in any of those measurements, or if it comes from some unknown physical phenomenon. For instance, some works have reported a slowly decreasing trend on $H_0$ with redshift using a binned analysis for different data \cite{Dainotti:2021pqg,Dainotti:2022bzg} and propose a $f(R)$ gravity could account for this variation \cite{Schiavone:2022wvq}. Others have considered the tension is alleviated by a phantom evolving dark energy \cite{Teng:2021cvy}, or could be solved assuming an underdensity known as the KBC void \cite{Haslbauer:2020xaa}, and yet some argue a breakdown of FLRW cosmology \cite{Krishnan:2021dyb}. A review of several proposed theoretical solutions can be found in \cite{DiValentino:2021izs}. In any case, additional and independent data could help us clarifying the origin of such disagreement.



Recently some works have proposed using the age of the oldest astrophysical objects (OAO) as an independent route to investigate the Hubble tension \cite{Jimenez:2019onw,Bernal:2021yli,Boylan-Kolchin:2021fvy,Krishnan:2021dyb,Vagnozzi:2021tjv,Wei:2022plg}. The ages of OAO played an important role establishing the $\Lambda$CDM model, with reports of OAO being older than the Universe assuming the prevailing Einstein-de Sitter model \cite{Dunlop:1996mp,Jimenez:1996at,VandenBerg:1996tm}. This led to an age crisis \cite{Jaffe:1995qu,1995Natur376399B,Krauss:1995yb,Ostriker:1995su,Alcaniz:1999kr}, which was solved by the discovery of the late-time cosmic acceleration and the necessity of a dark energy component using SNIa \cite{SupernovaSearchTeam:1998fmf,SupernovaCosmologyProject:1998vns}. After the end of the age crisis, the OAO received less attention, but some authors still used them to constrain dark energy and other cosmological parameters \cite{Jimenez:2003iv,Capozziello:2004jy,Samushia:2009px,Dantas:2010zh,Verde:2013fva,Bengaly:2013afa,Wei:2015cva,Rana:2016gha,Nunes:2020yij,Borghi:2021rft}. Now, recent data with more reliable and precise determination of the ages could provide additional constraints on the Hubble constant and enlighten the tension.

In this paper, we investigate the use of the ages of OAO to resolve the Hubble tension. Although, those data could in fact provide important information, we have found that the method can be biasing the results, artificially indicating a tension with other low-redshift data. Reanalyzing the ages of 114 OAO, we obtain that the constraint in the Hubble constant goes from slightly disfavoring local measurements, as in previous works, to favoring them.

The paper is organized as follows. In Section~\ref{Sec.2}, we will introduce two methods to constrain the Hubble function from the ages of the OAO used in the literature: estimating the minimum age of the Universe as a function of redshift and using lookback time. In Section~\ref{Sec.3}, we use 114 measurements of age-redshift data from OAO and re-obtain previous results in \cite{Vagnozzi:2021tjv,Wei:2022plg} to ensure our code produces the expected results. In Section~\ref{Sec.4}, we create a mock catalog of age-redshift values with known cosmology and test if those two methods recover the fiducial cosmology appropriately. Section~\ref{Sec.5} presents our cosmological constraints reanalyzing the real 114 age-redshift data correcting the likelihood for the age of the Universe. Finally, we summarize our conclusions in Section~\ref{Sec.6}.


\section{Methods using OAO}
\label{Sec.2}
\begin{figure}
\includegraphics[width=\columnwidth]{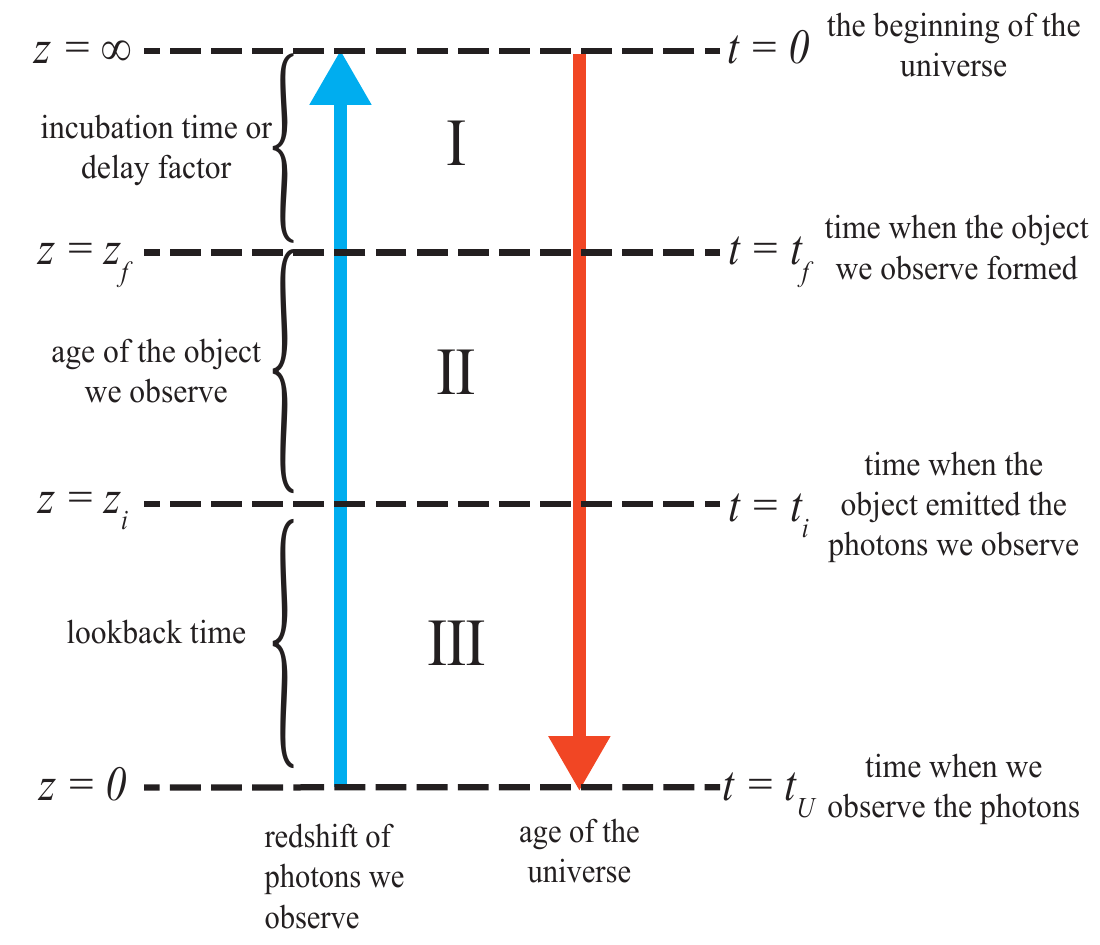}
\caption{Redshift-age evolution for the old astrophysical objects.}
\label{fig.1}
\end{figure}

As shown in Figure~\ref{fig.1}, we can separate the history of the Universe into three parts with respect to the evolution of the OAO we observe. Region~\uppercase\expandafter{\romannumeral1}, from the beginning of the Universe to the formation of the object, is what we call incubation time or delay factor. Region~\uppercase\expandafter{\romannumeral2} spans the evolution of the astrophysical object from its formation to the time they emit the photons we observe. These photons carry information about the age of those sources at the end of that period. At last, Region~\uppercase\expandafter{\romannumeral3} encompasses the time the photons spent on traveling from the time they were emitted until we detect them nowadays.

\subsection{Age of the Universe from the OAO}
The age of the OAO can be estimated via a wide variety of methods. The general methodology relies on fitting the spectral energy distributions with spectral population synthesis models, with further assumptions on dust attenuation, formation history, luminosity and colors. As the Universe must be at least as old as the objects it contains at any redshift, we can estimate a lower limit to the age of the Universe from the OAO \cite{Vagnozzi:2021tjv}.

The theoretical age of the Universe at redshift $z$ is determined by the age-redshift relation via the formula
\begin{equation}
\label{Eq.1}
t_{_U}(z) = \int^\infty_z{\frac{dz^{\prime}}{(1+z^{\prime})H(z^{\prime})}} \,,
\end{equation}
where $H(z)$ stands for the Hubble function and, in the $\Lambda$CDM model, $H(z)$ is given by
\begin{equation}
\label{Eq.2}
H(z) = H_0\sqrt{\Omega_m (1+z)^3+1-\Omega_m} \,.
\end{equation}
Therefore, requiring the Universe to be at least as old as the OAO imposes an upper limit to the value of the Hubble constant $H_0$.

Given $N$ observed data for the ages of OAO, $t^{obs}_i \pm \sigma_i$ at redshifts $z_i$, we can model the probability of a set of parameters $\mathbf{\Theta}$ via the half-Gaussian (log-)likelihood \citep{Vagnozzi:2021tjv}
\begin{equation}
\label{Eq.3}
\begin{split}
\chi^2_{OAO} &= -2 \ln{\mathcal{L}(\mathbf{\Theta}|data)}\\
&=\sum\limits_i^N\left \{
\begin{array}{ll}
\Delta_i^2(\mathbf{\Theta})/\sigma^2_{t^{obs}_i} & \text{ , if } \Delta_i(\mathbf{\Theta}) < 0 \\
0  & \text{ , if } \Delta_i(\mathbf{\Theta}) \geq 0 \,,
\end{array}
\right.
\end{split}
\end{equation}
where $\Delta_i(\mathbf{\Theta}) \equiv t_{_U}(\mathbf{\Theta}, z_i) - (t^{obs}_i+df)$ is the difference between the age of the Universe and the sum of the delay factor and the age of the $i$th OAO at redshift $z_i$. Equation~\ref{Eq.3} interprets the fact that: (1) parameters leading the age of the Universe to be younger than the age of the OAO plus the delay factor (i.e. $\Delta_i(\mathbf{\Theta}) < 0$) are (exponentially) unlikely, since the container should not be younger than its contents; (2) parameters making the Universe older than its contents are equally likely and hence cannot be distinguished from each other on the basis of age method alone.

\subsection{Lookback time measurement}
Figure~\ref{fig.1} shows us that the theoretical lookback time of the $i$th OAO can be calculated as
\begin{equation}
\label{Eq.4}
t_{_L}(z_i) = \int^{z_i}_0{\frac{dz^{\prime}}{(1+z^{\prime})H(z^{\prime})}} \,.    
\end{equation}
On the other hand, we can estimate the observational lookback time using the ages of OAO and age of the Universe inferred from observation as
\begin{equation}
\label{Eq.5}
\begin{array}{ll}
t^{obs}_{_L}(z_i) &= t^{obs}_{_U} - t_i  \\
     &= t^{obs}_{_U} - t^{obs}_{age}(z_i) - df,
\end{array}    
\end{equation}
where $t^{obs}_{age}(z_i)$ is the observed age of $i$th OAO and $df$ is the delay factor.

We assume the observed lookback time follows a Gaussian distribution around the theoretical value. Therefore, the (log-)likelihood is given by
\begin{equation}
\label{Eq.6}
\begin{split}
\chi^2_{LBT} &= -2 \ln{\mathcal{L}(\mathbf{\Theta}|data)}\\
&=\sum\limits_i^N\frac{[t_{_L}(z_i, \mathbf{\Theta})-t^{obs}_{_L}(z_i,df)]^2}{\sigma_i^2+\sigma_{_U}^2}+\frac{[t_{_U}(\mathbf{\Theta}) - t^{obs}_{U}]^2}{\sigma_{_U}^2},
\end{split}
\end{equation}
where $\sigma_i$ and $\sigma_{_U}$ stand for the errors of observations about ages of the $i$th OAO and the Universe, respectively. The delay factor is a nuisance parameter that is fitted simultaneously with the cosmological parameters. To find out the best estimate of parameters, we only need to maximize the likelihood function, or namely minimize the $\chi_{LBT}^2$ given by Eq.~\ref{Eq.6}.

\section{Estimating the Hubble constant from OAO}
\label{Sec.3}
\begin{figure}
\includegraphics[width=\columnwidth]{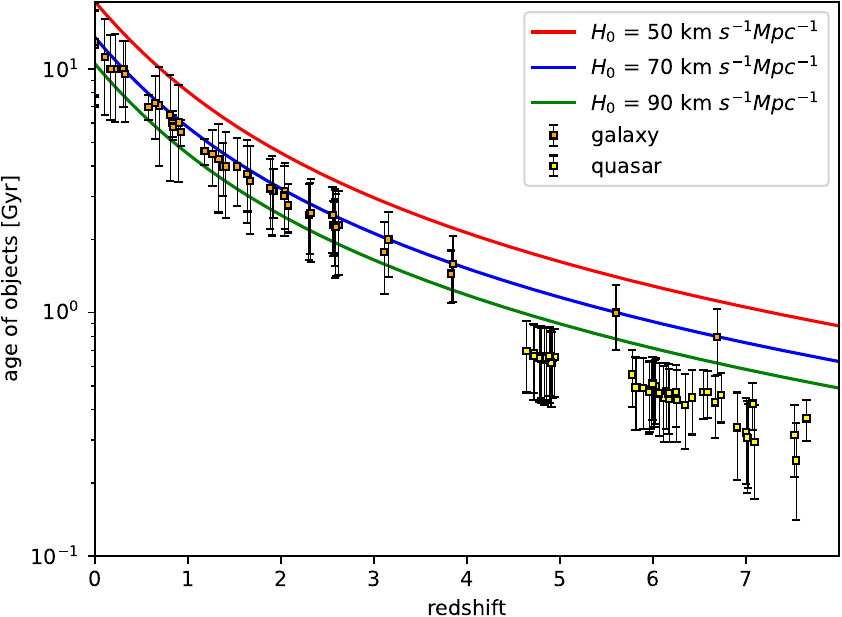}
\caption{Age-redshift relation for 114 OAO and also the theoretical estimation for the age of the Universe in a $\Lambda$CDM cosmology with $\Omega_m = 0.3$.}
\label{fig.2}
\end{figure}

In this section, we use real age-redshift data from 114 OAO and intend to re-obtain the same result for the Hubble constant as first shown in \cite{Vagnozzi:2021tjv} and later in \cite{Wei:2022plg}. The data is composed of the estimated ages of 61 galaxies and 53 quasars in the redshift range $z = [0, 8]$. Figure~\ref{fig.2} plot the age-redshift relation for these objects together with the theoretical estimation for the fiducial $\Lambda$CDM cosmology with $\Omega_m = 0.3$ and $H_0 = 70 \, \rm{km \, s^{-1} Mpc^{-1}}$, as well as for some variation in the Hubble constant.

In order to estimate an upper limit to the value of the Hubble constant, we use the age of the OAO and constrain the cosmological parameters with the likelihood described in Eq.~\ref{Eq.3}. As our model is composed by only 3 parameters, $\mathbf{\Theta} = \{H_0, \Omega_m, df \}$, we sample the parameter space using the brute force. Therefore, we calculate the likelihood for all values in a grid given by $H_0 \in [40,100] \SI{}{km \, s^{-1}Mpc^{-1}}$, $\Omega_m \in [0.2,0.4]$, and $df \in [0,1]\SI{}{Gyr}$ divided into smaller intervals with steps $\SI{0.01}{km \, s^{-1} Mpc^{-1}}$, $0.001$, and $\SI{0.01}{Gyr}$, respectively. Our posterior assumes a flat prior on $H_0$ and $\Omega_m$, and a J19 prior on $df$ (J19 prior comes from \cite{Jimenez:2019onw} and its fitting function is from \cite{Valcin:2020vav}) given by
\begin{equation}
\label{Eq.J19_1}
P(df) \propto 
0.95\exp{\left(-\frac{1}{2}\frac{(l-l_1)^2}{\sigma_1^2}\right)} + 0.45\exp{\left(-\frac{1}{2}\frac{(l-l_2)^2}{\sigma_2^2}\right)} \,,
\end{equation}
where $l=\log_{10}(df)$, $l_1\equiv\log_{10}(0.1155)$, and $l_2\equiv\log_{10}(0.255)$. In addition, $\sigma_2=0.155$ and $\sigma_1=0.15$, if $df\leq0.1155$, or $\sigma_1=0.17$, if $df > 0.1155$.

\begin{figure}
\includegraphics[width=\columnwidth]{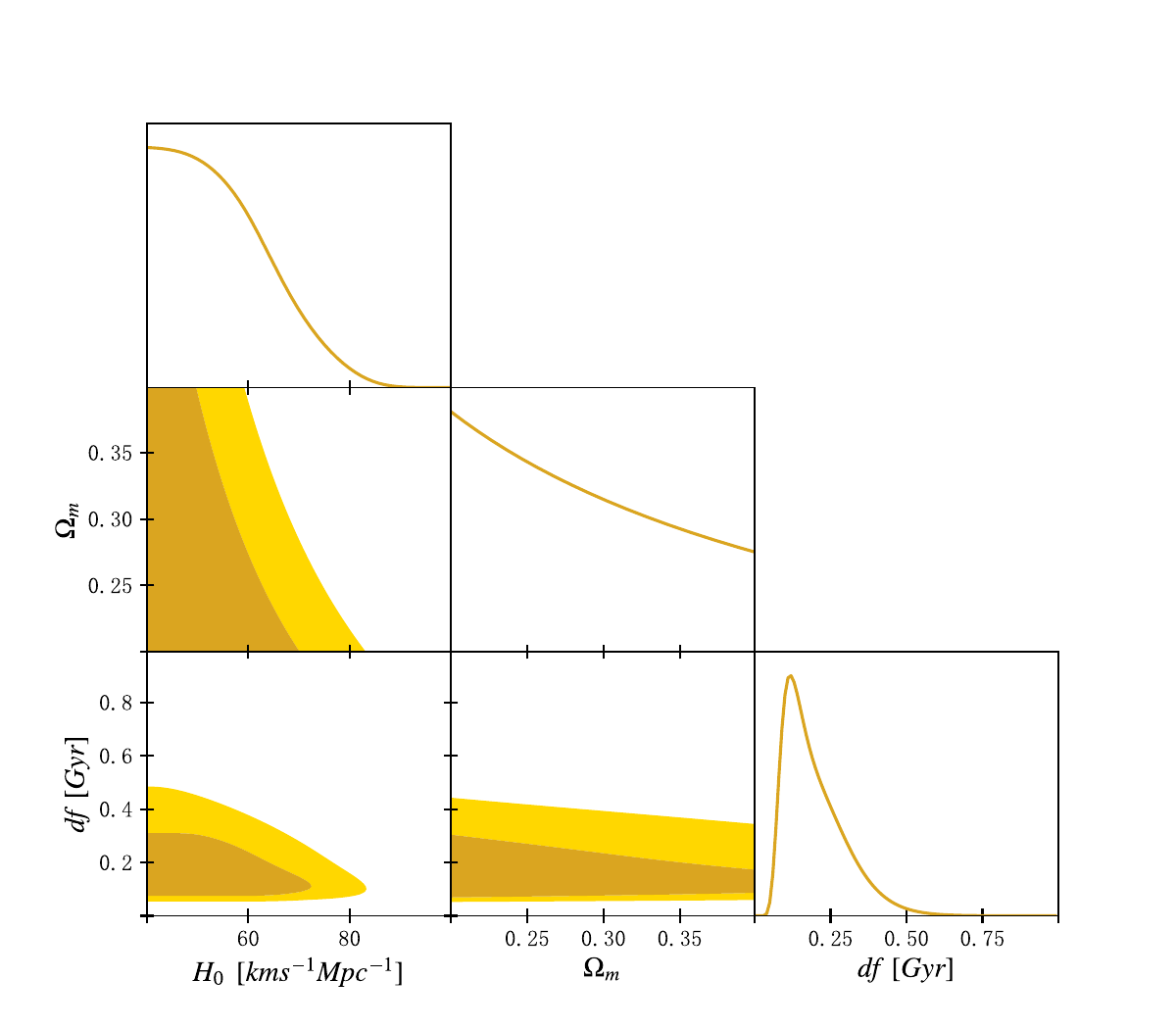}
\caption{2D and 1D marginalized posterior distribution with 68\% C.L. and 95\% C.L. for $H_0$, $\Omega_m$, and the delay factor $df$.}
\label{fig.3}
\end{figure}

We plot the 2D and 1D marginalized posterior distribution in Fig.~\ref{fig.3}. The 95\% C.L. upper limit on $H_0$ is $\SI{73.24}{km \, s^{-1} Mpc^{-1}}$ and the delay factor peaks at $df \approx \SI{0.15}{Gyr}$, which agree with the results in \cite{Vagnozzi:2021tjv} and \cite{Wei:2022plg}, although they used a Markov chain Monte Carlo (MCMC) sampling method.

\section{Simulated data}
\label{Sec.4}

We obtained in the previous section an upper limit on the value of $H_0$ using the assumption of half-Gaussian likelihood. Our value is consistent with previous ones, which makes us confident about our code and methodology. Vagnozzi et al. \cite{Vagnozzi:2021tjv} first proposed that the value obtained in this way is in slight disagreement with local measurements of the Hubble constant, in particular, the value of $H_0 = 74.03 \pm 1.42 \, \rm{km \, s^{-1} Mpc^{-1}}$ from the {\it Hubble Space Telescope} (HST) observations of 70 long-period Cepheids in the Large Magellanic Cloud \citep{Riess:2019cxk}. In this section, we intend to test these ideas and investigate the behaviour of the half-Gaussian likelihood presented in Eq.~\ref{Eq.3} in a full controlled scenario.

\begin{figure}
\includegraphics[width=\columnwidth]{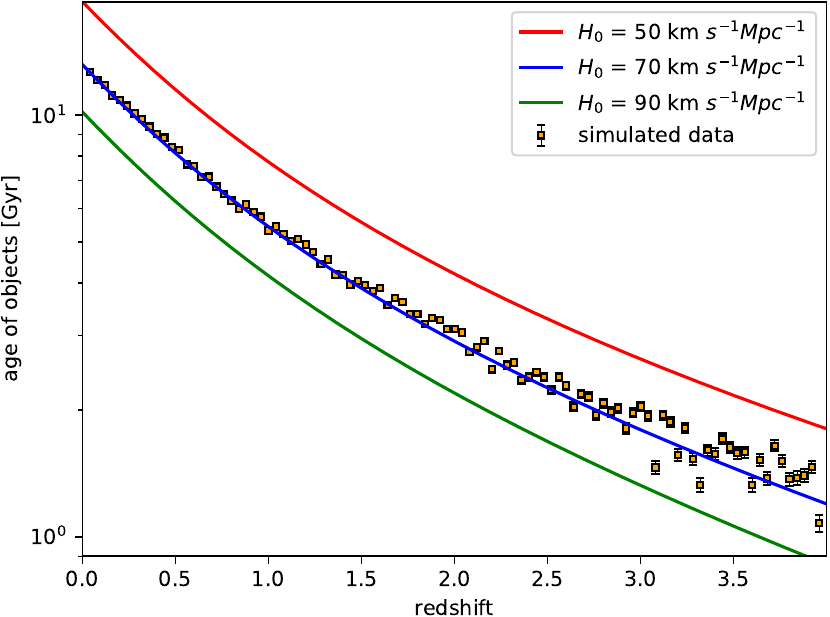}
\caption{Simulated age-redshift relation for 100 OAO with $\sigma_i = 0.05 \, \rm{Gyr}$. We also show the theoretical age-redshift relation in a $\Lambda$CDM cosmology with $\Omega_m = 0.3$ and some values for $H_0$. }
\label{fig.4}
\end{figure}

Assuming a fiducial $\Lambda$CDM cosmology with $H_0 = \SI{70.0}{km \, s^{-1} Mpc^{-1}}$ and $\Omega_m = 0.30$, we create a mock catalog of 100 age-redshift data. First, we obtain 100 uniformly distributed random points in the redshift range $z = (0, 4]$\footnote{A more realistic simulation for the distribution of OAO, such as galaxies or quasars, should properly take into account the redshift distribution of the sample, as well as instrumental and observational effects. The redshift distribution depends on the luminosity function, which is different for each specific sample as blue or red galaxies, and also evolves with redshift. In addition, observational and instrumental effects introduce cuts in the final distribution. However, any systematic effect in the age-redshift relation come from properly estimating the ages and redshifts of the sources. In our simple simulation, these uncertainties are modeled assuming a random Gaussian distribution. Therefore, we have more control and our results should be more consistent with the fiducial cosmology.}, then we calculate the age of the Universe at each redshift using Eq~\ref{Eq.1}. We subtract the age of the Universe at each redshift by a random delay factor $df$, which we obtain following the J19 distribution in Eq~\ref{Eq.J19_1}. Finally, we simulate the age of the OAO assuming a Gaussian distribution given by
\begin{equation}
\label{Eq.7}
P(t^{obs}_{i})=\frac{1}{\sqrt{2\pi}\sigma_i}\exp{\left(-\frac{[t^{obs}_{i} - (t_U(z_i) - df_i)]^2}{2\sigma_i^2}\right)} \,,
\end{equation}
where $\sigma_i$ is the error in the simulated data. For simplicity, we assume the error bars are the same for all objects. We consider two scenarios: a) $\sigma_i = 0.05 \, \rm{Gyr}$; and b) $\sigma = 1 \, \rm{Gyr}$. Figure~\ref{fig.4} shows our simulated data with $\sigma_i = 0.05 \, \rm{Gyr}$.

\begin{figure*}
\subfloat[]{
\includegraphics[width=0.47\textwidth]{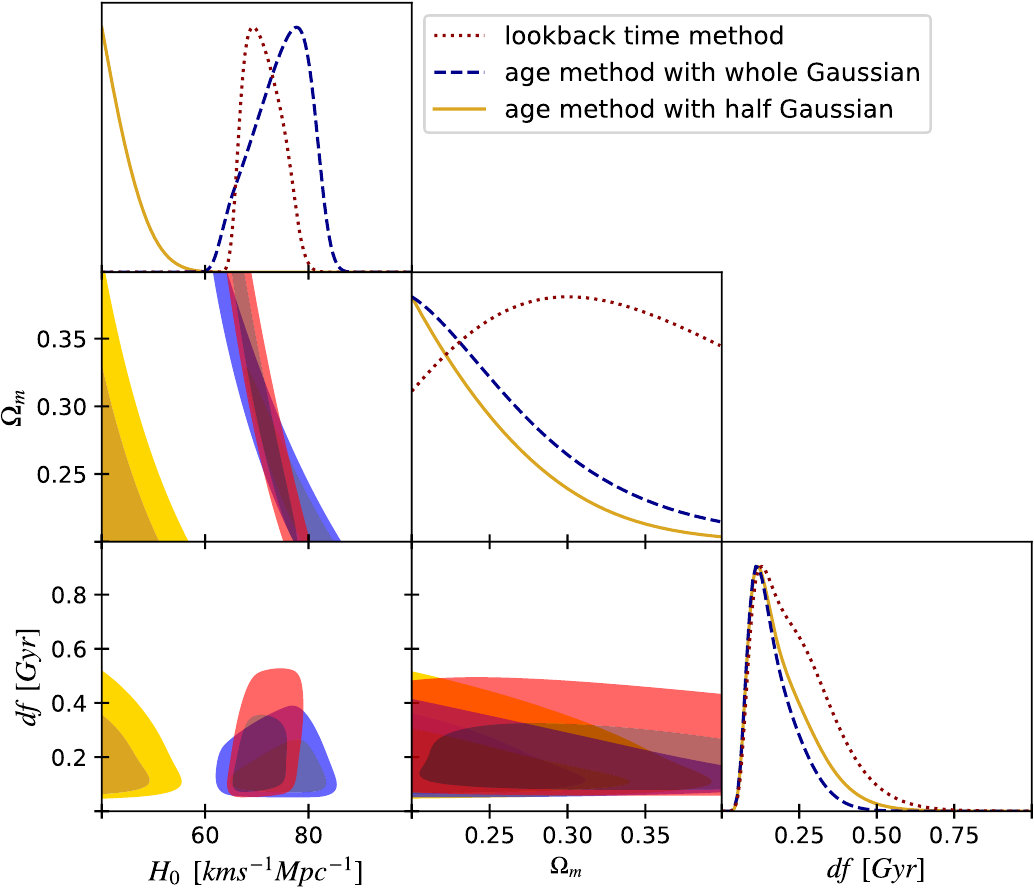}
}
\subfloat[]{
\includegraphics[width=0.47\textwidth]{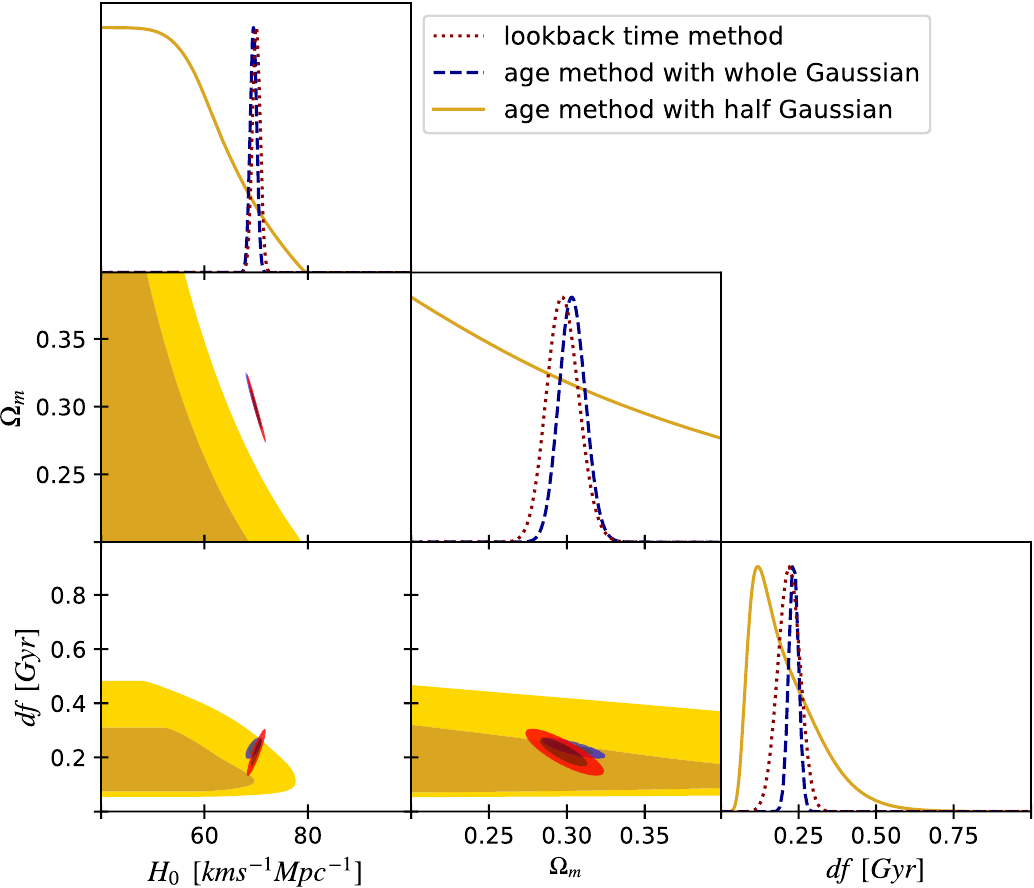}
}
\caption{2D and 1D marginalized posterior distribution using three different methods with the simulated data. We show the 68\% C.L. and 95\% C.L. for $H_0$, $\Omega_m$, and the delay factor $df$. (a) $\sigma_i = 1 \, \rm{Gyr}$ and (b) $\sigma_i = 0.05 \, \rm{Gyr}$.}
\label{fig.5}
\end{figure*}

We then redo the analysis made in Sect.~\ref{Sec.3} with the two mock catalogs. The 2D and 1D marginalized posterior distributions are shown in Fig.~\ref{fig.5} in solid yellow lines. By comparing our result with the previous one using the real data in Fig.~\ref{fig.3}, we observe they present the same behaviour. The 95\% C.L. upper limit on the Hubble constant is given by $H_0 \leq 70.84 \, \rm{km \, s^{-1}Mpc^{-1}}$ (for $\sigma_i = 0.05 \, \rm{Gyr}$) and $H_0 \leq 51.25 \, \rm{km \, s^{-1}Mpc^{-1}}$ (for $\sigma_i = 1 \, \rm{Gyr}$). As we can see, although the fiducial value in the simulation was set as $H_0 = 70 \, \rm{km \, s^{-1}Mpc^{-1}}$, the larger the error bars, the upper limit becomes less consistent with the fiducial value. This indicates that this method to constrain our cosmological parameters is not appropriate and is artificially imposing some tension with the correct value of $H_0$.

In order to further investigate this phenomenon, we redo the analysis but assuming the whole Gaussian in Eq.~\ref{Eq.3} (i.e. $\chi^2_{OAO} \sim \Delta_i^2(\mathbf{\Theta})/\sigma^2_{t^{obs}_i}$ for all values). Our results are plotted in Fig.~\ref{fig.5} in dashed blue lines. As expected, the constraints are consistent with the fiducial values used in our simulation, with marginalized average and 68\% C.L. given by $H_0 = 69.5 \pm 0.64 \, \rm{km \, s^{-1} Mpc^{-1}}$, $\Omega_m = 0.3041 \pm  0.0085$, and $df = 0.234 \pm 0.016 \, \rm{Gyr}$ (for $\sigma_i = 0.05 \, \rm{Gyr}$); and $H_0 = 74.58 \pm 5.19 \, \rm{km \, s^{-1} Mpc^{-1}}$, $\Omega_m = 0.2621 \pm 0.0484 $, and $df = 0.169 \pm 0.077 \, \rm{Gyr}$ (for $\sigma_i = 1 \, \rm{Gyr}$). We also observe that the delay factor is dominated by the prior distribution J19 in our posterior for large error bars as $\sigma_i = 1 \, \rm{Gyr}$, but presents a Gaussian distribution for small error bars. This comes from the fact that although our simulation uses the J19 distribution as expected in the real data, our likelihood actually models the delay factor as a constant for all OAO. Therefore, we only recover the J19 shape when the prior is dominating our data. This is also the case when we use the real data as in Fig.~\ref{fig.3}.

Similarly, we also repeat the analysis using the lookback time likelihood described in Eq.~\ref{Eq.6}. In this case, in addition to our simulated OAO data at redshifts $z_i$, we need the age of the Universe today. For simplicity, we simulate the age of the Universe with the same error as for the astrophysical objects $\sigma_U = \sigma_i$. The constraints using the lookback time are shown in Fig.~\ref{fig.5} in dotted red lines. We observe they are consistent with the whole Gaussian result and recover the correct values used in the simulation with marginalized constraints given by $H_0 = 70.04 \pm 0.75 \, \rm{km \, s^{-1} Mpc^{-1}}$, $\Omega_m = 0.2980 \pm 0.0103$, and $df = 0.219 \pm 0.035 \, \rm{Gyr}$ (for $\sigma_i = 0.05 \, \rm{Gyr}$); and $H_0 = 71.32 \pm 3.35 \, \rm{km \, s^{-1} Mpc^{-1}}$, $\Omega_m = 0.3019 \pm 0.0551$, and $df = 0.240 \pm 0.122 \, \rm{Gyr}$ (for $\sigma_i = 1 \, \rm{Gyr}$).

\section{Reanalyzing the OAO data}
\label{Sec.5}

\begin{figure}
\includegraphics[width=\columnwidth]{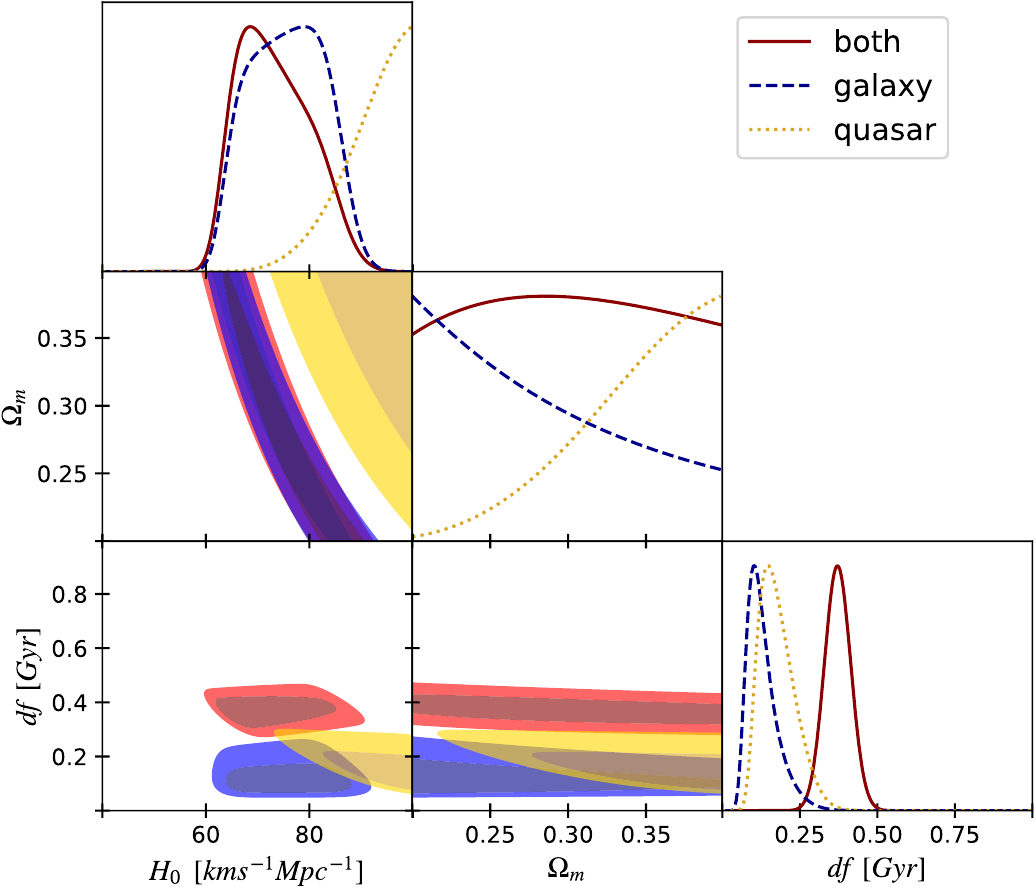}
\caption{2D and 1D marginalized posterior distribution using the age method with the real data and a likelihood assuming the whole Gaussian. We show the 68\% C.L. and 95\% C.L. for $H_0$, $\Omega_m$, and the delay factor $df$. }
\label{fig.7}
\end{figure}

In the previous section, we discovered that our assumption about the likelihood of the ages of OAO data can artificially introduce a tension between the inferred value for the Hubble constant and the actual one. Therefore, we now reanalyze the real data in Sect.~\ref{Sec.3} using the whole Gaussian likelihood. Our constraints are presented in Fig.~\ref{fig.7}. We consider both the 114 OAO data together and also the constraints coming from galaxies or quasars separately. Figure~\ref{fig.2} already indicates that the data from galaxies and quasars are not consistent with each other, and we actually observe this in the plot $\Omega_m \times H_0$ of Fig.~\ref{fig.7}. The origin for this inconsistency is not clear for us yet, but unless it can be fixed, these two data sets should be treated separately\footnote{Risaliti and Russo also studied the cosmological constraints from the Hubble diagram of quasars in the redshift range $0.5 < z < 5.5$ and reported that the distance modulus-redshift relation of quasars at $z < 1.4$ is in agreement with that of supernovae and with the concordance model. However, a deviation from the $\Lambda$CDM model emerges at higher redshifts, with a statistical significance of $\sim 4\sigma$ \cite{Risaliti:2018reu}.}.

Figure~\ref{fig.7} also tells us that the constraints from the ages of the OAO are still weak. The data, especially for quasars, tend to $\Omega_m > 0.4$. However, we fix our prior to $\Omega_m \in [0.2, 0.4]$, which is consistent with previous works and is also large enough to contain any realistic cosmology. Therefore, our results move the posterior from slightly disfavoring local measurements as in Sect.~\ref{Sec.3} to favoring them. The marginalized average and 68\% C.L. values we obtain are $H_0 = 75.79 \pm 7.05 \, \rm{km \, s^{-1} Mpc^{-1}}$, $\Omega_m = 0.2791 \pm 0.0558$, and $df = 0.131 \pm 0.051 \, \rm{Gyr}$ (only galaxies); $H_0 = 91.67 \pm 6.30 \, \rm{km \, s^{-1} Mpc^{-1}}$, $\Omega_m = 0.3416 \pm 0.0426$, and $df = 0.179 \pm 0.057 \, \rm{Gyr}$ (only quasars); and $H_0 = 73.44 \pm 6.89 \, \rm{km \, s^{-1} Mpc^{-1}}$, $\Omega_m = 0.2994 \pm 0.0566$, and $df = 0.372 \pm 0.042 \, \rm{Gyr}$ (for both).

\section{Conclusions}
\label{Sec.6}
The tension in the Hubble constant between the value measured from low redshift data and inferred from the CMB data is a major problem in cosmology today. It is still not clear if this tension originates from systematic effects in our data or from new physics. Therefore, the use of new and independent data is crucial to understand this phenomenon. It has then been proposed that the ages of the OAO can help shedding light in this matter.

Although it is true that the OAO provide an independent source to study this tension, we show that the methodology can also artificially introduce some tension. We discuss how this can happen and be influencing some previous works, and propose another way to analyze the ages of the OAO. We test those methods using a full controlled simulation and also reanalyze real age-redshift data.

The ages of the OAO with a whole Gaussian likelihood produces results similar to the lookback time if the uncertainty is small. But the later also depends on the age of the Universe, which is used as an anchor for all lookback time measurements. Therefore, we prefer to consider the first case only in order to obtain results about the Hubble tension. Our constraints are still weak, but favors low redshift measurements in contrast to the {\it Planck} result.

Finally, we also show that the ages of quasars as in Fig.~\ref{fig.2} are not consistent with the ages of galaxies. It is not possible, in the $\Lambda$CDM model, to find parameters that are simultaneously compatible with both of them. It is not clear for us yet the origin of such disagreement, but it seems to be related with the calibration of their ages.


\backmatter

\bmhead{Acknowledgments}

We would like to express our sincere thanks to Dr. Jun-Jie Wei for sharing the compilation of the age-redshift data from OAO with us. We also thank Prof. Adam Riess for helpful comments in the first version of this manuscript. A.A.C also acknowledges financial support from the National Natural Science Foundation of China (grant 12175192).

\bibliography{references}


\begin{thebibliography}{38}
\ifx \bisbn   \undefined \def \bisbn  #1{ISBN #1}\fi
\ifx \binits  \undefined \def \binits#1{#1}\fi
\ifx \bauthor  \undefined \def \bauthor#1{#1}\fi
\ifx \batitle  \undefined \def \batitle#1{#1}\fi
\ifx \bjtitle  \undefined \def \bjtitle#1{#1}\fi
\ifx \bvolume  \undefined \def \bvolume#1{\textbf{#1}}\fi
\ifx \byear  \undefined \def \byear#1{#1}\fi
\ifx \bissue  \undefined \def \bissue#1{#1}\fi
\ifx \bfpage  \undefined \def \bfpage#1{#1}\fi
\ifx \blpage  \undefined \def \blpage #1{#1}\fi
\ifx \burl  \undefined \def \burl#1{\textsf{#1}}\fi
\ifx \doiurl  \undefined \def \doiurl#1{\url{https://doi.org/#1}}\fi
\ifx \betal  \undefined \def \betal{\textit{et al.}}\fi
\ifx \binstitute  \undefined \def \binstitute#1{#1}\fi
\ifx \binstitutionaled  \undefined \def \binstitutionaled#1{#1}\fi
\ifx \bctitle  \undefined \def \bctitle#1{#1}\fi
\ifx \beditor  \undefined \def \beditor#1{#1}\fi
\ifx \bpublisher  \undefined \def \bpublisher#1{#1}\fi
\ifx \bbtitle  \undefined \def \bbtitle#1{#1}\fi
\ifx \bedition  \undefined \def \bedition#1{#1}\fi
\ifx \bseriesno  \undefined \def \bseriesno#1{#1}\fi
\ifx \blocation  \undefined \def \blocation#1{#1}\fi
\ifx \bsertitle  \undefined \def \bsertitle#1{#1}\fi
\ifx \bsnm \undefined \def \bsnm#1{#1}\fi
\ifx \bsuffix \undefined \def \bsuffix#1{#1}\fi
\ifx \bparticle \undefined \def \bparticle#1{#1}\fi
\ifx \barticle \undefined \def \barticle#1{#1}\fi
\bibcommenthead
\ifx \bconfdate \undefined \def \bconfdate #1{#1}\fi
\ifx \botherref \undefined \def \botherref #1{#1}\fi
\ifx \url \undefined \def \url#1{\textsf{#1}}\fi
\ifx \bchapter \undefined \def \bchapter#1{#1}\fi
\ifx \bbook \undefined \def \bbook#1{#1}\fi
\ifx \bcomment \undefined \def \bcomment#1{#1}\fi
\ifx \oauthor \undefined \def \oauthor#1{#1}\fi
\ifx \citeauthoryear \undefined \def \citeauthoryear#1{#1}\fi
\ifx \endbibitem  \undefined \def \endbibitem {}\fi
\ifx \bconflocation  \undefined \def \bconflocation#1{#1}\fi
\ifx \arxivurl  \undefined \def \arxivurl#1{\textsf{#1}}\fi
\csname PreBibitemsHook\endcsname

\bibitem[\protect\citeauthoryear{Aghanim et~al.}{2020a}]{Planck:2018nkj}
\begin{barticle}
\bauthor{\bsnm{Aghanim}, \binits{N.}}, \betal:
\batitle{{Planck 2018 results. I. Overview and the cosmological legacy of
  Planck}}.
\bjtitle{Astron. Astrophys.}
\bvolume{641},
\bfpage{1}
(\byear{2020})
\doiurl{10.1051/0004-6361/201833880}
{\href{https://arxiv.org/abs/1807.06205}{{arXiv:1807.06205}}}
{[astro-ph.CO]}
\end{barticle}
\endbibitem

\bibitem[\protect\citeauthoryear{Aghanim et~al.}{2020b}]{Planck:2018vyg}
\begin{barticle}
\bauthor{\bsnm{Aghanim}, \binits{N.}}, \betal:
\batitle{{Planck 2018 results. VI. Cosmological parameters}}.
\bjtitle{Astron. Astrophys.}
\bvolume{641},
\bfpage{6}
(\byear{2020})
\doiurl{10.1051/0004-6361/201833910}
{\href{https://arxiv.org/abs/1807.06209}{{arXiv:1807.06209}}}
{[astro-ph.CO]}.
\bcomment{[Erratum: Astron.Astrophys. 652, C4 (2021)]}
\end{barticle}
\endbibitem

\bibitem[\protect\citeauthoryear{Riess et~al.}{2022}]{Riess:2021jrx}
\begin{barticle}
\bauthor{\bsnm{Riess}, \binits{A.G.}}, \betal:
\batitle{{A Comprehensive Measurement of the Local Value of the Hubble Constant
  with 1 km s$^{-1}$ Mpc$^{-1}$ Uncertainty from the Hubble Space Telescope and
  the SH0ES Team}}.
\bjtitle{Astrophys. J. Lett.}
\bvolume{934}(\bissue{1}),
\bfpage{7}
(\byear{2022})
\doiurl{10.3847/2041-8213/ac5c5b}
{\href{https://arxiv.org/abs/2112.04510}{{arXiv:2112.04510}}}
{[astro-ph.CO]}
\end{barticle}
\endbibitem

\bibitem[\protect\citeauthoryear{Dainotti et~al.}{2021}]{Dainotti:2021pqg}
\begin{barticle}
\bauthor{\bsnm{Dainotti}, \binits{M.G.}},
\bauthor{\bsnm{De~Simone}, \binits{B.}},
\bauthor{\bsnm{Schiavone}, \binits{T.}},
\bauthor{\bsnm{Montani}, \binits{G.}},
\bauthor{\bsnm{Rinaldi}, \binits{E.}},
\bauthor{\bsnm{Lambiase}, \binits{G.}}:
\batitle{{On the Hubble constant tension in the SNe Ia Pantheon sample}}.
\bjtitle{Astrophys. J.}
\bvolume{912}(\bissue{2}),
\bfpage{150}
(\byear{2021})
\doiurl{10.3847/1538-4357/abeb73}
{\href{https://arxiv.org/abs/2103.02117}{{arXiv:2103.02117}}}
{[astro-ph.CO]}
\end{barticle}
\endbibitem

\bibitem[\protect\citeauthoryear{Dainotti et~al.}{2022}]{Dainotti:2022bzg}
\begin{barticle}
\bauthor{\bsnm{Dainotti}, \binits{M.G.}},
\bauthor{\bsnm{De~Simone}, \binits{B.}},
\bauthor{\bsnm{Schiavone}, \binits{T.}},
\bauthor{\bsnm{Montani}, \binits{G.}},
\bauthor{\bsnm{Rinaldi}, \binits{E.}},
\bauthor{\bsnm{Lambiase}, \binits{G.}},
\bauthor{\bsnm{Bogdan}, \binits{M.}},
\bauthor{\bsnm{Ugale}, \binits{S.}}:
\batitle{{On the Evolution of the Hubble Constant with the SNe Ia Pantheon
  Sample and Baryon Acoustic Oscillations: A Feasibility Study for
  GRB-Cosmology in 2030}}.
\bjtitle{Galaxies}
\bvolume{10}(\bissue{1}),
\bfpage{24}
(\byear{2022})
\doiurl{10.3390/galaxies10010024}
{\href{https://arxiv.org/abs/2201.09848}{{arXiv:2201.09848}}}
{[astro-ph.CO]}
\end{barticle}
\endbibitem

\bibitem[\protect\citeauthoryear{Schiavone et~al.}{2023}]{Schiavone:2022wvq}
\begin{barticle}
\bauthor{\bsnm{Schiavone}, \binits{T.}},
\bauthor{\bsnm{Montani}, \binits{G.}},
\bauthor{\bsnm{Bombacigno}, \binits{F.}}:
\batitle{{f(R) gravity in the Jordan frame as a paradigm for the Hubble
  tension}}.
\bjtitle{Mon. Not. Roy. Astron. Soc.}
\bvolume{522}(\bissue{1}),
\bfpage{72}--\blpage{77}
(\byear{2023})
\doiurl{10.1093/mnrasl/slad041}
{\href{https://arxiv.org/abs/2211.16737}{{arXiv:2211.16737}}}
{[gr-qc]}
\end{barticle}
\endbibitem

\bibitem[\protect\citeauthoryear{Teng et~al.}{2021}]{Teng:2021cvy}
\begin{barticle}
\bauthor{\bsnm{Teng}, \binits{Y.-P.}},
\bauthor{\bsnm{Lee}, \binits{W.}},
\bauthor{\bsnm{Ng}, \binits{K.-W.}}:
\batitle{{Constraining the dark-energy equation of state with cosmological
  data}}.
\bjtitle{Phys. Rev. D}
\bvolume{104}(\bissue{8}),
\bfpage{083519}
(\byear{2021})
\doiurl{10.1103/PhysRevD.104.083519}
{\href{https://arxiv.org/abs/2105.02667}{{arXiv:2105.02667}}}
{[astro-ph.CO]}
\end{barticle}
\endbibitem

\bibitem[\protect\citeauthoryear{Haslbauer et~al.}{2020}]{Haslbauer:2020xaa}
\begin{barticle}
\bauthor{\bsnm{Haslbauer}, \binits{M.}},
\bauthor{\bsnm{Banik}, \binits{I.}},
\bauthor{\bsnm{Kroupa}, \binits{P.}}:
\batitle{{The KBC void and Hubble tension contradict $\Lambda$CDM on a Gpc
  scale \ensuremath{-} Milgromian dynamics as a possible solution}}.
\bjtitle{Mon. Not. Roy. Astron. Soc.}
\bvolume{499}(\bissue{2}),
\bfpage{2845}--\blpage{2883}
(\byear{2020})
\doiurl{10.1093/mnras/staa2348}
{\href{https://arxiv.org/abs/2009.11292}{{arXiv:2009.11292}}}
{[astro-ph.CO]}
\end{barticle}
\endbibitem

\bibitem[\protect\citeauthoryear{Krishnan et~al.}{2021}]{Krishnan:2021dyb}
\begin{barticle}
\bauthor{\bsnm{Krishnan}, \binits{C.}},
\bauthor{\bsnm{Mohayaee}, \binits{R.}},
\bauthor{\bsnm{Colg\'ain}, \binits{E.O.}},
\bauthor{\bsnm{Sheikh-Jabbari}, \binits{M.M.}},
\bauthor{\bsnm{Yin}, \binits{L.}}:
\batitle{{Does Hubble tension signal a breakdown in FLRW cosmology?}}
\bjtitle{Class. Quant. Grav.}
\bvolume{38}(\bissue{18}),
\bfpage{184001}
(\byear{2021})
\doiurl{10.1088/1361-6382/ac1a81}
{\href{https://arxiv.org/abs/2105.09790}{{arXiv:2105.09790}}}
{[astro-ph.CO]}
\end{barticle}
\endbibitem

\bibitem[\protect\citeauthoryear{Di~Valentino
  et~al.}{2021}]{DiValentino:2021izs}
\begin{barticle}
\bauthor{\bsnm{Di~Valentino}, \binits{E.}},
\bauthor{\bsnm{Mena}, \binits{O.}},
\bauthor{\bsnm{Pan}, \binits{S.}},
\bauthor{\bsnm{Visinelli}, \binits{L.}},
\bauthor{\bsnm{Yang}, \binits{W.}},
\bauthor{\bsnm{Melchiorri}, \binits{A.}},
\bauthor{\bsnm{Mota}, \binits{D.F.}},
\bauthor{\bsnm{Riess}, \binits{A.G.}},
\bauthor{\bsnm{Silk}, \binits{J.}}:
\batitle{{In the realm of the Hubble tension\textemdash{}a review of
  solutions}}.
\bjtitle{Class. Quant. Grav.}
\bvolume{38}(\bissue{15}),
\bfpage{153001}
(\byear{2021})
\doiurl{10.1088/1361-6382/ac086d}
{\href{https://arxiv.org/abs/2103.01183}{{arXiv:2103.01183}}}
{[astro-ph.CO]}
\end{barticle}
\endbibitem

\bibitem[\protect\citeauthoryear{Jimenez et~al.}{2019}]{Jimenez:2019onw}
\begin{barticle}
\bauthor{\bsnm{Jimenez}, \binits{R.}},
\bauthor{\bsnm{Cimatti}, \binits{A.}},
\bauthor{\bsnm{Verde}, \binits{L.}},
\bauthor{\bsnm{Moresco}, \binits{M.}},
\bauthor{\bsnm{Wandelt}, \binits{B.}}:
\batitle{{The local and distant Universe: stellar ages and $H_0$}}.
\bjtitle{JCAP}
\bvolume{03},
\bfpage{043}
(\byear{2019})
\doiurl{10.1088/1475-7516/2019/03/043}
{\href{https://arxiv.org/abs/1902.07081}{{arXiv:1902.07081}}}
{[astro-ph.CO]}
\end{barticle}
\endbibitem

\bibitem[\protect\citeauthoryear{Bernal et~al.}{2021}]{Bernal:2021yli}
\begin{barticle}
\bauthor{\bsnm{Bernal}, \binits{J.L.}},
\bauthor{\bsnm{Verde}, \binits{L.}},
\bauthor{\bsnm{Jimenez}, \binits{R.}},
\bauthor{\bsnm{Kamionkowski}, \binits{M.}},
\bauthor{\bsnm{Valcin}, \binits{D.}},
\bauthor{\bsnm{Wandelt}, \binits{B.D.}}:
\batitle{{The trouble beyond $H_0$ and the new cosmic triangles}}.
\bjtitle{Phys. Rev. D}
\bvolume{103}(\bissue{10}),
\bfpage{103533}
(\byear{2021})
\doiurl{10.1103/PhysRevD.103.103533}
{\href{https://arxiv.org/abs/2102.05066}{{arXiv:2102.05066}}}
{[astro-ph.CO]}
\end{barticle}
\endbibitem

\bibitem[\protect\citeauthoryear{Boylan-Kolchin and
  Weisz}{2021}]{Boylan-Kolchin:2021fvy}
\begin{barticle}
\bauthor{\bsnm{Boylan-Kolchin}, \binits{M.}},
\bauthor{\bsnm{Weisz}, \binits{D.R.}}:
\batitle{{Uncertain times: the redshift\textendash{}time relation from
  cosmology and stars}}.
\bjtitle{Mon. Not. Roy. Astron. Soc.}
\bvolume{505}(\bissue{2}),
\bfpage{2764}--\blpage{2783}
(\byear{2021})
\doiurl{10.1093/mnras/stab1521}
{\href{https://arxiv.org/abs/2103.15825}{{arXiv:2103.15825}}}
{[astro-ph.CO]}
\end{barticle}
\endbibitem

\bibitem[\protect\citeauthoryear{Vagnozzi et~al.}{2022}]{Vagnozzi:2021tjv}
\begin{barticle}
\bauthor{\bsnm{Vagnozzi}, \binits{S.}},
\bauthor{\bsnm{Pacucci}, \binits{F.}},
\bauthor{\bsnm{Loeb}, \binits{A.}}:
\batitle{{Implications for the Hubble tension from the ages of the oldest
  astrophysical objects}}.
\bjtitle{JHEAp}
\bvolume{36},
\bfpage{27}--\blpage{35}
(\byear{2022})
\doiurl{10.1016/j.jheap.2022.07.004}
{\href{https://arxiv.org/abs/2105.10421}{{arXiv:2105.10421}}}
{[astro-ph.CO]}
\end{barticle}
\endbibitem

\bibitem[\protect\citeauthoryear{Wei and Melia}{2022}]{Wei:2022plg}
\begin{barticle}
\bauthor{\bsnm{Wei}, \binits{J.-J.}},
\bauthor{\bsnm{Melia}, \binits{F.}}:
\batitle{{Exploring the Hubble Tension and Spatial Curvature from the Ages of
  Old Astrophysical Objects}}.
\bjtitle{Astrophys. J.}
\bvolume{928}(\bissue{2}),
\bfpage{165}
(\byear{2022})
\doiurl{10.3847/1538-4357/ac562c}
{\href{https://arxiv.org/abs/2202.07865}{{arXiv:2202.07865}}}
{[astro-ph.CO]}
\end{barticle}
\endbibitem

\bibitem[\protect\citeauthoryear{Dunlop et~al.}{1996}]{Dunlop:1996mp}
\begin{barticle}
\bauthor{\bsnm{Dunlop}, \binits{J.}},
\bauthor{\bsnm{Peacock}, \binits{J.}},
\bauthor{\bsnm{Spinrad}, \binits{H.}},
\bauthor{\bsnm{Dey}, \binits{A.}},
\bauthor{\bsnm{Jimenez}, \binits{R.}},
\bauthor{\bsnm{Stern}, \binits{D.}},
\bauthor{\bsnm{Windhorst}, \binits{R.}}:
\batitle{{A 3.5 - Gyr - old galaxy at redshift 1.55.}}
\bjtitle{Nature}
\bvolume{381},
\bfpage{581}
(\byear{1996})
\doiurl{10.1038/381581a0}
\end{barticle}
\endbibitem

\bibitem[\protect\citeauthoryear{Jimenez et~al.}{1996}]{Jimenez:1996at}
\begin{barticle}
\bauthor{\bsnm{Jimenez}, \binits{R.}},
\bauthor{\bsnm{Thejll}, \binits{P.}},
\bauthor{\bsnm{Jorgensen}, \binits{U.}},
\bauthor{\bsnm{MacDonald}, \binits{J.}},
\bauthor{\bsnm{Pagel}, \binits{B.}}:
\batitle{{Ages of globular clusters: a new approach}}.
\bjtitle{Mon. Not. Roy. Astron. Soc.}
\bvolume{282},
\bfpage{926}--\blpage{942}
(\byear{1996})
\doiurl{10.1093/mnras/282.3.926}
{\href{https://arxiv.org/abs/astro-ph/9602132}{{arXiv:astro-ph/9602132}}}
\end{barticle}
\endbibitem

\bibitem[\protect\citeauthoryear{VandenBerg et~al.}{1996}]{VandenBerg:1996tm}
\begin{barticle}
\bauthor{\bsnm{VandenBerg}, \binits{D.A.}},
\bauthor{\bsnm{Bolte}, \binits{M.}},
\bauthor{\bsnm{Stetson}, \binits{P.B.}}:
\batitle{{The age of the galactic globular cluster system}}.
\bjtitle{Ann. Rev. Astron. Astrophys.}
\bvolume{34},
\bfpage{461}--\blpage{510}
(\byear{1996})
\doiurl{10.1146/annurev.astro.34.1.461}
\end{barticle}
\endbibitem

\bibitem[\protect\citeauthoryear{Jaffe}{1996}]{Jaffe:1995qu}
\begin{barticle}
\bauthor{\bsnm{Jaffe}, \binits{A.H.}}:
\batitle{{H0 and odds on cosmology}}.
\bjtitle{Astrophys. J.}
\bvolume{471},
\bfpage{24}
(\byear{1996})
\doiurl{10.1086/177950}
{\href{https://arxiv.org/abs/astro-ph/9501070}{{arXiv:astro-ph/9501070}}}
\end{barticle}
\endbibitem

\bibitem[\protect\citeauthoryear{{Bolte} and {Hogan}}{1995}]{1995Natur376399B}
\begin{barticle}
\bauthor{\bsnm{{Bolte}}, \binits{M.}},
\bauthor{\bsnm{{Hogan}}, \binits{C.J.}}:
\batitle{{Conflict over the age of the Universe}}.
\bjtitle{Nature}
\bvolume{376}(\bissue{6539}),
\bfpage{399}--\blpage{402}
(\byear{1995})
\doiurl{10.1038/376399a0}
\end{barticle}
\endbibitem

\bibitem[\protect\citeauthoryear{Krauss and Turner}{1995}]{Krauss:1995yb}
\begin{barticle}
\bauthor{\bsnm{Krauss}, \binits{L.M.}},
\bauthor{\bsnm{Turner}, \binits{M.S.}}:
\batitle{{The Cosmological constant is back}}.
\bjtitle{Gen. Rel. Grav.}
\bvolume{27},
\bfpage{1137}--\blpage{1144}
(\byear{1995})
\doiurl{10.1007/BF02108229}
{\href{https://arxiv.org/abs/astro-ph/9504003}{{arXiv:astro-ph/9504003}}}
\end{barticle}
\endbibitem

\bibitem[\protect\citeauthoryear{Ostriker and
  Steinhardt}{1995}]{Ostriker:1995su}
\begin{barticle}
\bauthor{\bsnm{Ostriker}, \binits{J.P.}},
\bauthor{\bsnm{Steinhardt}, \binits{P.J.}}:
\batitle{{The Observational case for a low density universe with a nonzero
  cosmological constant}}.
\bjtitle{Nature}
\bvolume{377},
\bfpage{600}--\blpage{602}
(\byear{1995})
\doiurl{10.1038/377600a0}
\end{barticle}
\endbibitem

\bibitem[\protect\citeauthoryear{Alcaniz and Lima}{1999}]{Alcaniz:1999kr}
\begin{barticle}
\bauthor{\bsnm{Alcaniz}, \binits{J.S.}},
\bauthor{\bsnm{Lima}, \binits{J.A.S.}}:
\batitle{{New limits on omega\_ lambda and omega\_m from old galaxies at high
  redshift}}.
\bjtitle{Astrophys. J. Lett.}
\bvolume{521},
\bfpage{87}
(\byear{1999})
\doiurl{10.1086/312191}
{\href{https://arxiv.org/abs/astro-ph/9902298}{{arXiv:astro-ph/9902298}}}
\end{barticle}
\endbibitem

\bibitem[\protect\citeauthoryear{Riess
  et~al.}{1998}]{SupernovaSearchTeam:1998fmf}
\begin{barticle}
\bauthor{\bsnm{Riess}, \binits{A.G.}}, \betal:
\batitle{{Observational evidence from supernovae for an accelerating universe
  and a cosmological constant}}.
\bjtitle{Astron. J.}
\bvolume{116},
\bfpage{1009}--\blpage{1038}
(\byear{1998})
\doiurl{10.1086/300499}
{\href{https://arxiv.org/abs/astro-ph/9805201}{{arXiv:astro-ph/9805201}}}
\end{barticle}
\endbibitem

\bibitem[\protect\citeauthoryear{Perlmutter
  et~al.}{1999}]{SupernovaCosmologyProject:1998vns}
\begin{barticle}
\bauthor{\bsnm{Perlmutter}, \binits{S.}}, \betal:
\batitle{{Measurements of $\Omega$ and $\Lambda$ from 42 high redshift
  supernovae}}.
\bjtitle{Astrophys. J.}
\bvolume{517},
\bfpage{565}--\blpage{586}
(\byear{1999})
\doiurl{10.1086/307221}
{\href{https://arxiv.org/abs/astro-ph/9812133}{{arXiv:astro-ph/9812133}}}
\end{barticle}
\endbibitem

\bibitem[\protect\citeauthoryear{Jimenez et~al.}{2003}]{Jimenez:2003iv}
\begin{barticle}
\bauthor{\bsnm{Jimenez}, \binits{R.}},
\bauthor{\bsnm{Verde}, \binits{L.}},
\bauthor{\bsnm{Treu}, \binits{T.}},
\bauthor{\bsnm{Stern}, \binits{D.}}:
\batitle{{Constraints on the equation of state of dark energy and the Hubble
  constant from stellar ages and the CMB}}.
\bjtitle{Astrophys. J.}
\bvolume{593},
\bfpage{622}--\blpage{629}
(\byear{2003})
\doiurl{10.1086/376595}
{\href{https://arxiv.org/abs/astro-ph/0302560}{{arXiv:astro-ph/0302560}}}
\end{barticle}
\endbibitem

\bibitem[\protect\citeauthoryear{Capozziello et~al.}{2004}]{Capozziello:2004jy}
\begin{barticle}
\bauthor{\bsnm{Capozziello}, \binits{S.}},
\bauthor{\bsnm{Cardone}, \binits{V.F.}},
\bauthor{\bsnm{Funaro}, \binits{M.}},
\bauthor{\bsnm{Andreon}, \binits{S.}}:
\batitle{{Constraining dark energy models using the lookback time to galaxy
  clusters and the age of the Universe}}.
\bjtitle{Phys. Rev. D}
\bvolume{70},
\bfpage{123501}
(\byear{2004})
\doiurl{10.1103/PhysRevD.70.123501}
{\href{https://arxiv.org/abs/astro-ph/0410268}{{arXiv:astro-ph/0410268}}}
\end{barticle}
\endbibitem

\bibitem[\protect\citeauthoryear{Samushia et~al.}{2010}]{Samushia:2009px}
\begin{barticle}
\bauthor{\bsnm{Samushia}, \binits{L.}},
\bauthor{\bsnm{Dev}, \binits{A.}},
\bauthor{\bsnm{Jain}, \binits{D.}},
\bauthor{\bsnm{Ratra}, \binits{B.}}:
\batitle{{Constraints on dark energy from the lookback time versus redshift
  test}}.
\bjtitle{Phys. Lett. B}
\bvolume{693},
\bfpage{509}--\blpage{514}
(\byear{2010})
\doiurl{10.1016/j.physletb.2010.07.057}
{\href{https://arxiv.org/abs/0906.2734}{{arXiv:0906.2734}}}
{[astro-ph.CO]}
\end{barticle}
\endbibitem

\bibitem[\protect\citeauthoryear{Dantas et~al.}{2011}]{Dantas:2010zh}
\begin{barticle}
\bauthor{\bsnm{Dantas}, \binits{M.A.}},
\bauthor{\bsnm{Alcaniz}, \binits{J.S.}},
\bauthor{\bsnm{Mania}, \binits{D.}},
\bauthor{\bsnm{Ratra}, \binits{B.}}:
\batitle{{Distance and time constraints on accelerating cosmological models}}.
\bjtitle{Phys. Lett. B}
\bvolume{699},
\bfpage{239}--\blpage{245}
(\byear{2011})
\doiurl{10.1016/j.physletb.2011.04.014}
{\href{https://arxiv.org/abs/1010.0995}{{arXiv:1010.0995}}}
{[astro-ph.CO]}
\end{barticle}
\endbibitem

\bibitem[\protect\citeauthoryear{Verde et~al.}{2013}]{Verde:2013fva}
\begin{barticle}
\bauthor{\bsnm{Verde}, \binits{L.}},
\bauthor{\bsnm{Jimenez}, \binits{R.}},
\bauthor{\bsnm{Feeney}, \binits{S.}}:
\batitle{{The importance of local measurements for cosmology}}.
\bjtitle{Phys. Dark Univ.}
\bvolume{2},
\bfpage{65}--\blpage{71}
(\byear{2013})
\doiurl{10.1016/j.dark.2013.04.003}
{\href{https://arxiv.org/abs/1303.5341}{{arXiv:1303.5341}}}
{[astro-ph.CO]}
\end{barticle}
\endbibitem

\bibitem[\protect\citeauthoryear{Bengaly et~al.}{2014}]{Bengaly:2013afa}
\begin{barticle}
\bauthor{\bsnm{Bengaly}, \binits{C.A.P.} \bsuffix{Jr.}},
\bauthor{\bsnm{Dantas}, \binits{M.A.}},
\bauthor{\bsnm{Carvalho}, \binits{J.C.}},
\bauthor{\bsnm{Alcaniz}, \binits{J.S.}}:
\batitle{{Forecasting cosmological constraints from age of high-z galaxies}}.
\bjtitle{Astron. Astrophys.}
\bvolume{561},
\bfpage{44}
(\byear{2014})
\doiurl{10.1051/0004-6361/201322475}
{\href{https://arxiv.org/abs/1308.6230}{{arXiv:1308.6230}}}
{[astro-ph.CO]}
\end{barticle}
\endbibitem

\bibitem[\protect\citeauthoryear{Wei et~al.}{2015}]{Wei:2015cva}
\begin{barticle}
\bauthor{\bsnm{Wei}, \binits{J.-J.}},
\bauthor{\bsnm{Wu}, \binits{X.-F.}},
\bauthor{\bsnm{Melia}, \binits{F.}},
\bauthor{\bsnm{Wang}, \binits{F.-Y.}},
\bauthor{\bsnm{Yu}, \binits{H.}}:
\batitle{{The Age-Redshift Relationship of Old Passive Galaxies}}.
\bjtitle{Astron. J.}
\bvolume{150},
\bfpage{35}
(\byear{2015})
\doiurl{10.1088/0004-6256/150/1/35}
{\href{https://arxiv.org/abs/1505.07671}{{arXiv:1505.07671}}}
{[astro-ph.CO]}
\end{barticle}
\endbibitem

\bibitem[\protect\citeauthoryear{Rana et~al.}{2017}]{Rana:2016gha}
\begin{barticle}
\bauthor{\bsnm{Rana}, \binits{A.}},
\bauthor{\bsnm{Jain}, \binits{D.}},
\bauthor{\bsnm{Mahajan}, \binits{S.}},
\bauthor{\bsnm{Mukherjee}, \binits{A.}}:
\batitle{{Constraining cosmic curvature by using age of galaxies and
  gravitational lenses}}.
\bjtitle{JCAP}
\bvolume{03},
\bfpage{028}
(\byear{2017})
\doiurl{10.1088/1475-7516/2017/03/028}
{\href{https://arxiv.org/abs/1611.07196}{{arXiv:1611.07196}}}
{[astro-ph.CO]}
\end{barticle}
\endbibitem

\bibitem[\protect\citeauthoryear{Nunes and Pacucci}{2020}]{Nunes:2020yij}
\begin{barticle}
\bauthor{\bsnm{Nunes}, \binits{R.C.}},
\bauthor{\bsnm{Pacucci}, \binits{F.}}:
\batitle{{Effects of the Hubble Parameter on the Cosmic Growth of the First
  Quasars}}.
\bjtitle{Mon. Not. Roy. Astron. Soc.}
\bvolume{496}(\bissue{1}),
\bfpage{888}--\blpage{893}
(\byear{2020})
\doiurl{10.1093/mnras/staa1568}
{\href{https://arxiv.org/abs/2006.01839}{{arXiv:2006.01839}}}
{[astro-ph.CO]}
\end{barticle}
\endbibitem

\bibitem[\protect\citeauthoryear{Borghi et~al.}{2022}]{Borghi:2021rft}
\begin{barticle}
\bauthor{\bsnm{Borghi}, \binits{N.}},
\bauthor{\bsnm{Moresco}, \binits{M.}},
\bauthor{\bsnm{Cimatti}, \binits{A.}}:
\batitle{{Toward a Better Understanding of Cosmic Chronometers: A New
  Measurement of H(z) at z \ensuremath{\sim} 0.7}}.
\bjtitle{Astrophys. J. Lett.}
\bvolume{928}(\bissue{1}),
\bfpage{4}
(\byear{2022})
\doiurl{10.3847/2041-8213/ac3fb2}
{\href{https://arxiv.org/abs/2110.04304}{{arXiv:2110.04304}}}
{[astro-ph.CO]}
\end{barticle}
\endbibitem

\bibitem[\protect\citeauthoryear{Valcin et~al.}{2020}]{Valcin:2020vav}
\begin{barticle}
\bauthor{\bsnm{Valcin}, \binits{D.}},
\bauthor{\bsnm{Bernal}, \binits{J.L.}},
\bauthor{\bsnm{Jimenez}, \binits{R.}},
\bauthor{\bsnm{Verde}, \binits{L.}},
\bauthor{\bsnm{Wandelt}, \binits{B.D.}}:
\batitle{{Inferring the Age of the Universe with Globular Clusters}}.
\bjtitle{JCAP}
\bvolume{12},
\bfpage{002}
(\byear{2020})
\doiurl{10.1088/1475-7516/2020/12/002}
{\href{https://arxiv.org/abs/2007.06594}{{arXiv:2007.06594}}}
{[astro-ph.CO]}
\end{barticle}
\endbibitem

\bibitem[\protect\citeauthoryear{Riess et~al.}{2019}]{Riess:2019cxk}
\begin{barticle}
\bauthor{\bsnm{Riess}, \binits{A.G.}},
\bauthor{\bsnm{Casertano}, \binits{S.}},
\bauthor{\bsnm{Yuan}, \binits{W.}},
\bauthor{\bsnm{Macri}, \binits{L.M.}},
\bauthor{\bsnm{Scolnic}, \binits{D.}}:
\batitle{{Large Magellanic Cloud Cepheid Standards Provide a 1\% Foundation for
  the Determination of the Hubble Constant and Stronger Evidence for Physics
  beyond $\Lambda$CDM}}.
\bjtitle{Astrophys. J.}
\bvolume{876}(\bissue{1}),
\bfpage{85}
(\byear{2019})
\doiurl{10.3847/1538-4357/ab1422}
{\href{https://arxiv.org/abs/1903.07603}{{arXiv:1903.07603}}}
{[astro-ph.CO]}
\end{barticle}
\endbibitem

\bibitem[\protect\citeauthoryear{Risaliti and Lusso}{2019}]{Risaliti:2018reu}
\begin{barticle}
\bauthor{\bsnm{Risaliti}, \binits{G.}},
\bauthor{\bsnm{Lusso}, \binits{E.}}:
\batitle{{Cosmological constraints from the Hubble diagram of quasars at high
  redshifts}}.
\bjtitle{Nature Astron.}
\bvolume{3}(\bissue{3}),
\bfpage{272}--\blpage{277}
(\byear{2019})
\doiurl{10.1038/s41550-018-0657-z}
{\href{https://arxiv.org/abs/1811.02590}{{arXiv:1811.02590}}}
{[astro-ph.CO]}
\end{barticle}
\endbibitem

\end{thebibliography}

\end{document}